\documentstyle[12pt,aaspp4]{article}

\begin{document}

\title{Neutrino Fluence after $r$-Process Freeze-Out and Abundances of Te
Isotopes in Presolar Diamonds}
\author{Y.-Z. Qian and P. Vogel}
\affil{Physics Department, California Institute of Technology,
       Pasadena, CA 91125}
\authoremail{yzqian@citnp.caltech.edu, vogel@lamppost.caltech.edu}
\and
\author{G. J. Wasserburg}
\affil{The Lunatic Asylum,
Division of Geological and Planetary Sciences, California
       Institute of Technology, Pasadena, CA 91125}

\begin{abstract}

Using the data of Richter et al. (1998) on Te isotopes in diamond grains from
a meteorite, we derive bounds on the neutrino fluence and the decay
timescale of the neutrino flux relevant for the supernova $r$-process.
Our new bound on the neutrino fluence ${\cal{F}}$ after freeze-out of
the $r$-process peak at mass number $A\sim 130$ is more stringent than
the previous bound ${\cal{F}}\lesssim 0.045$ 
(in units of $10^{37}$~erg~cm$^{-2}$)
of Qian et al. (1997) and Haxton et al. (1997) if the neutrino flux decays
on a timescale $\hat\tau\gtrsim 0.65$~s. In particular, it requires that a
fluence of ${\cal{F}}=0.031$ be provided by a neutrino flux with
$\hat\tau\lesssim 0.84$~s. 
Such a fluence may be responsible for the production
of the solar $r$-process abundances at $A=124$--126 (Qian et al. 1997;
Haxton et al. 1997). Our results are based on the assumption that
only the stable nuclei implanted into the diamonds are retained while the
radioactive ones are lost from the diamonds upon decay after implantation
(Ott 1996). We consider that the nanodiamonds
are condensed in an environment with ${\rm C}/{\rm O}>1$ in the expanding 
supernova debris or from the exterior H envelope. This environment need
not have the $^{13}$C/$^{12}$C ratio of the bulk diamonds as 
the Te- and Xe-containing nanodiamond grains are 
too rare to affect that ratio.
The implantation of nuclei would have occurred
$\sim 10^{4}$--$10^{6}$~s after $r$-process freeze-out. 
This time interval may be marginally sufficient to permit
adequate cooling upon expansion for the formation of diamond grains. 
The mechanisms of
preferential retention/loss of the implanted nuclei are not well understood.

\end{abstract}


\section{Introduction}

The nature and occurrence of presolar diamonds and the associated
isotopic anomalies in several elements have been reported on and discussed
by various authors (see e.g., Fraundorf et al. 1989; Russell, Arden, \&
Pillinger 1991; Anders \& Zinner 1993; Huss \& Lewis 1995). 
It has been shown that macroscopic samples of these diamond crystals 
($\sim 1$~nm in size) contain xenon with major excesses of
$^{124}$Xe, $^{126}$Xe, $^{134}$Xe and $^{136}$Xe. This xenon has a
characteristic abundance pattern which is called Xe-HL. These excesses
appear to be correlated and have been attributed to a variety of causes
ranging from super heavy element fission (Anders et al. 1975) 
to some special combination
of $r$-process and $p$-process. It is evident that this
widespread anomalous Xe (cf. Huss \& Lewis, 1995) cannot be attributed to
a single nuclear process and could represent a mixture of two or more
distinctive nucleosynthetic sources that were subsequently mixed. The
separation of Xe-H (with excesses in $^{134}$Xe and $^{136}$Xe
of a likely $r$-process origin) from Xe-L
(with excesses in $^{124}$Xe and $^{126}$Xe of a likely $p$-process origin) 
appears to have been accomplished by Meshik,
Pravdivtseva, \& Hohenberg (1998) in an elegant experiment. This
demonstrates that Xe-H is a distinct component.

Recently, Richter, Ott, \& Begemann (1998) have shown that tellurium found
in diamond-rich residues from a meteorite is comprised of a mixture of
isotopically normal Te (e.g., terrestrial composition) and a pure
$r$-process component of $^{128}$Te and $^{130}$Te. However, $^{125}$Te and
$^{126}$Te with substantial $r$-process contributions are essentially
absent in the component containing $^{128}$Te and $^{130}$Te. Earlier,
based on the Xe data and the fact that stable isotopes are produced through
$\beta$-decays after $r$-process freeze-out, Ott (1996) proposed that the
isotopic composition of the pure $r$-process component in presolar diamonds
reflects the lifetimes of precursors for individual isotopes. He also
calculated a period of $\sim 7.5\times 10^3$~s 
between $r$-process freeze-out and
implantation of $r$-process products into the diamonds, assuming that only
the stable isotopes present at the time of implantation will be retained.
The study by Richter et al. (1998) strongly supports his proposal, as the
virtual absence of pure $r$-process $^{125}$Te and $^{126}$Te in the
diamonds can be attributed to the long half-lives of their corresponding
precursors $^{125}$Sb (2.76 yr) and $^{126}$Sn ( $\sim 10^5$~yr). In the
model of Ott (1996), it is required that radioactive nuclei will be lost
almost completely from the diamonds upon decay after implantation.

In this paper we discuss the connection between the isotopic composition
of presolar diamonds and the supernova $r$-process. In particular, we
consider the effects of a neutrino flux on the distribution of Te isotopes
after $r$-process freeze-out. The high abundance of $^{128}$Sn, the
precursor for $^{128}$Te, could facilitate appreciable production of
$^{126}$Te due to interactions with neutrinos. Using the data of Richter et
al. (1998), we can derive a bound on the neutrino fluence to which nuclei
were subject subsequent to the freeze-out of the $r$-process in a supernova
environment. This bound is presented in \S2. We recognize that the
mechanism of implantation and retention/loss of volatile elements like Te
and Xe in the very fine diamond crystallites is still
enigmatic. These and other general issues will be discussed later in \S3.
For simplicity, we assume the following time sequence and environment
illustrated in Figure~1. 

At the time $t_{\rm ej}$, some material is ejected from the neutron star
produced by the core collapse and the subsequent supernova explosion.
Within the moving material, the $r$-process starts at $t_0$ and continues
to $t_{\rm FO}$, the time of freeze-out. By definition, the $r$-process
involves rapid neutron capture onto seed nuclei (with $\beta$-decays and
possibly also $\nu_e$ capture reactions playing important roles, see, e.g.,
Kratz et al. 1993; Fuller \& Meyer 1995; McLaughlin \& Fuller 1996, 1997;
Qian et al. 1997; Haxton et al. 1997; Qian, Vogel, \& Wasserburg 1998). At
$t_{\rm FO}$, which is taken to coincide with the time $t=0$, all the
neutrons are exhausted and the population of nuclei is assumed to reach the
$r$-process peak at the magic neutron number $N=82$, where the abundances
of all nuclear species would correspond to the normal solar $r$-process
yields at mass number $A\sim 130$ if these nuclei had time to $\beta$-decay.
Subsequent to freeze-out (i.e., at $t>0$), these nuclei are subject to
modification by further neutrino interactions 
(for $\sim 10$~s, the so-called ``postprocessing'') and
$\beta$-decays and to implantation into the pre-existing diamond grains
(with retention of only the stable nuclei).

Diamond grains are assumed to be formed at the time $t_{\rm DF}$ after
freeze-out and nuclei can be implanted into them anytime after. 
At $t_{\rm ret}$ ($\sim$ a few hours),
the grains are sufficiently cool so that the
implanted stable nuclei will be retained indefinitely. However, radioactive
nuclei will, by recoil or by lattice site damage, be lost from the
microcrystals upon decay after implantation (Ott 1996). We now address the
effects of neutrino-nucleus interactions immediately after $t_{\rm FO}$
that can significantly alter the abundances of stable Te isotopes 
at $t_{\rm ret}$,
and derive a bound on the neutrino fluence after $r$-process freeze-out.

\section{Bound on Neutrino Fluence after $r$-Process Freeze-Out}

The Te isotopic abundance ratios are 
${^{125}{\rm Te}}:{^{126}{\rm Te}}:{^{128}{\rm Te}}:{^{130}{\rm Te}}
= 0.0006 \pm 0.0018:0.0001 \pm 0.0037:1.008 \pm 0.018:1$
for the pure $r$-process component in presolar
diamonds (Richter, Ott, \& Begemann 1998), compared with
$0.159:0.307:0.923:1$ for the solar $r$-process component (K\"appeler,
Beer, \& Wisshak 1989). This striking difference can be explained by the
hypothesis of Ott (1996) that the diamonds sample only the $r$-process
yields carried by stable isotopes a few hours after freeze-out. With a
generic precursor abundance pattern at freeze-out, the solar $r$-process
abundances of Te isotopes would be produced after all $\beta$-decays are
completed. The rather close agreement 
between the $^{128}$Te/$^{130}$Te ratio in
the diamonds and that in the solar system indicates that $^{128}$Te and
$^{130}$Te are essentially carrying their full $r$-process yields a few
hours after freeze-out, as their precursors 
have half-lives of $\lesssim 1$~hr.
On the other hand, the virtual absence of $^{125}$Te and $^{126}$Te
in the diamonds appears to be due to the extremely slow $\beta$-decays of
their corresponding precursors $^{125}$Sb and $^{126}$Sn.

Although essentially no production of $^{126}$Te occurs through the
$\beta$-decay of $^{126}$Sn on a timescale of a few hours, $^{126}$Te can
be produced on a much shorter timescale via neutrino reactions after the
$r$-process freezes out in a supernova environment. These reactions are:
${^{128}{\rm Sn}}(\nu_e,e^-2n){^{126}{\rm Sb}}$, 
${^{127}{\rm Sn}}(\nu_e,e^-n){^{126}{\rm Sb}}$, and 
${^{126}{\rm Sn}}(\nu_e,e^-){^{126}{\rm Sb}}$, where, for example, 
${^{128}{\rm Sn}}(\nu_e,e^-2n){^{126}{\rm Sb}}$ 
is a short-hand for $\nu_e$ capture on
$^{128}$Sn producing $^{126}$Sb, an electron, and two neutrons. Reactions
with emission of more than two neutrons are insignificant due to energetic
requirements. The above reactions of concern
produce $^{126}$Sb mainly in its isomeric state, 
which has a half-life of 19.15~minutes and $\beta$-decays to
$^{126}$Te 86\% of the time. (The ground state of $^{126}$Sb has spin 8 and
is unlikely to be populated directly by neutrino interactions due to
angular momentum selection rules.) Therefore, a bound on the neutrino
fluence after $r$-process freeze-out can be obtained by restricting
neutrino-induced production of $^{126}$Te. However, no similar bound can be
obtained by considering $^{125}$Te. While $\nu_e$ capture reactions on Sb
isotopes can bypass the slow $\beta$-decay of $^{125}$Sb (which holds up the
$^{125}$Te production), no neutrino flux would be available to induce such
reactions. This is because the 
$\beta$-decays only reach the relevant Sb isotopes more than
1~minute after $r$-process freeze-out while the entire supernova neutrino
emission only lasts for $\sim 10$~s.

Compared with $^{126}$Sn and $^{127}$Sn,
$^{128}$Sn is more abundant and is reached earlier when the
neutrino flux is also higher after $r$-process freeze-out. 
Therefore, while the cross sections for neutrino reactions on $^{126}$Sn,
$^{127}$Sn, and $^{128}$Sn are similar,
the reaction on $^{128}$Sn is the most
important one for the production of $^{126}$Te.
In what follows, we specifically consider the reaction
${^{128}{\rm Sn}}(\nu_e,e^-2n){^{126}{\rm Sb}}$ 
to derive a bound on the neutrino fluence after freeze-out.
We separate the process of neutrino-induced
production of $^{126}$Te into three steps: (a) the production of $^{128}$Sn
through $\beta$-decays after $r$-process freeze-out, (b) the production of
$^{126}$Sb by ${^{128}{\rm Sn}}(\nu_e,e^-2n){^{126}{\rm Sb}}$, and (c) the
production of $^{126}$Te through the $\beta$-decay of $^{126}$Sb. These
three steps are illustrated in Figure~2.

The production of $^{128}$Sn is essentially through a chain of
$\beta$-decays starting from $^{128}$Pd (with $N=82$) at $r$-process
freeze-out. For simplicity, we do not consider neutrino reactions at this
stage, although adding them would speed up the production of $^{128}$Sn
somewhat and might give a more stringent bound on the neutrino fluence. The
half-life of $^{128}$Sn is 59.07~minutes and it can be taken as stable over
a period of $\sim 10$~s relevant for neutrino-induced production of
$^{126}$Sb. The other half-lives in the decay chain are 0.125, 0.0922,
0.34, and 0.84~s for $^{128}$Pd, $^{128}$Ag, $^{128}$Cd, and $^{128}$In,
respectively. The first two half-lives are taken from M\"oller, Nix, \&
Kratz (1997), and the other two from Firestone et al. (1996).

The $\nu_e$ capture on $^{128}$Sn mainly proceeds through Fermi and
Gamow-Teller (GT) transitions. The Fermi strength is carried by the
isobaric analog state (IAS) in $^{128}$Sb, which lies below the two-neutron
emission threshold. Guided by experimental data on $^{128}$Te and
$^{130}$Te (Madey et al. 1989), we put 85\% of the GT strength in the giant
resonance (GTGR) states and the rest in the low-lying states. The cross
section for ${^{128}{\rm Sn}}(\nu_e,e^-2n){^{126}{\rm Sb}}$ 
is essentially determined
by transitions to the GTGR states between the thresholds for emission of
two and three neutrons. The spectrum-averaged cross section for this
reaction is 
$\langle\sigma_{\nu_e,2n}\rangle=1.65\times 10^{-41}\ {\rm cm}^2$,
where the $\nu_e$ spectrum is taken to be
$f_{\nu_e}(E_{\nu_e})\propto E_{\nu_e}^2/[\exp(E_{\nu_e}/T_{\nu_e}-3)+1]$
with an average energy $\langle E_{\nu_e}\rangle=3.99\,T_{\nu_e}=11$~MeV
(see e.g., Janka \& Hillebrandt 1989). The corresponding reaction rate is 
\begin{equation}
\lambda_{\nu_e,2n}(t)={L_{\nu_e}(t)\over 4\pi r(t)^2}
{\langle\sigma_{\nu_e,2n}\rangle\over\langle E_{\nu_e}\rangle}
=0.744{L_{\nu_e,51}(t)\over r_7(t)^2}\ {\rm s}^{-1},
\end{equation}
where $L_{\nu_e,51}$ is the $\nu_e$ luminosity $L_{\nu_e}$ in units of
$10^{51}$~erg~s$^{-1}$, and $r_7$ is the distance $r$ of the $r$-process
material from the neutron star in units of $10^7$~cm. After $r$-process
freeze-out (i.e., at $t>0$), we assume that 
\begin{equation}
{L_{\nu_e,51}(t)\over r_7(t)^2}=
{L_{\nu_e,51}(0)\over r_7(0)^2}\exp\left(-{t\over\hat\tau}\right),
\end{equation}
where $\hat\tau$ is the characteristic decay timescale of the neutrino flux.

As mentioned earlier, 
the reaction ${^{128}{\rm Sn}}(\nu_e,e^-2n){^{126}{\rm Sb}}$
produces $^{126}$Sb mainly in its isomeric state,
which directly $\beta$-decays to $^{126}$Te 86\% of the time.
We do not consider neutrino reactions on $^{126}$Sb in the production of
$^{126}$Te as their effects are of the second order in the neutrino fluence. 

In order to illuminate the underlying physics, we first derive a bound on
the neutrino fluence after $r$-process freeze-out using a perturbative
approach and assuming that all of the four $\beta$-decays leading to
$^{128}$Sn have the same rate of $\bar\lambda_\beta=1.98\ {\rm s}^{-1}$.
This rate corresponds to a lifetime that equals a quarter of the sum of
lifetimes in the original decay chain. Without any neutrino effects,
the abundance of $^{128}$Sn after freeze-out is 
\begin{equation}
Y_{^{128}{\rm Sn}}(t)=1-\sum_{n=0}^3{(\bar\lambda_\beta t)^n\over n!}
\exp(-\bar\lambda_\beta t),
\label{y}
\end{equation}
where we have set the abundance of $^{128}$Pd at freeze-out
($t=0$) to unity without loss
of generality. The neutrino-induced production
of $^{126}$Sb over a period of $\sim 10$~s after freeze-out 
can be evaluated by the first order perturbation theory as
\begin{equation}
(\Delta Y)_{^{126}{\rm Sb}}=
\int_0^\infty\lambda_{\nu_e,2n}(t)Y_{^{128}{\rm Sn}}(t)dt
={0.744{\cal{F}}\over[1+(\bar\lambda_\beta\hat\tau)^{-1}]^4},
\label{dy}
\end{equation}
where 
\begin{equation}
{\cal{F}}=\int_0^\infty{L_{\nu_e,51}(t)\over r_7(t)^2}dt
={L_{\nu_e,51}(0)\over r_7(0)^2}\hat\tau
\end{equation}
is the neutrino fluence after freeze-out in units of $10^{37}$~erg~cm$^{-2}$.
For practical purpose, we have taken $\sim 10$~s as infinity in 
equation~(\ref{dy}). 

A few hours after $r$-process freeze-out, 86\% of the $^{126}$Sb produced
by neutrino reactions has decayed to $^{126}$Te. At the same time, the bulk
of the initial $^{128}$Pd has decayed to $^{128}$Te. From the abundance
ratio ${^{126}{\rm Te}}/{^{128}{\rm Te}}<3.67\times 10^{-3}$ in presolar
diamonds, we obtain 
\begin{equation}
{\cal{F}}<{3.67\times 10^{-3}\over 0.744\times 0.86}
\left(1+{1\over\bar\lambda_\beta\hat\tau}\right)^4=
5.74\times 10^{-3}\left[1+\left({0.505\ {\rm s}\over\hat\tau}\right)\right]^4.
\label{f}
\end{equation}
The bound given in equation~(\ref{f}) is shown as the dot-dashed line
in Figure~3.

Using the actual $\beta$-decay rates for the production of $^{128}$Sn and
taking into account its destruction by $\nu_e$ capture with all neutron
emission channels and by inelastic neutral-current scatterings of
$\nu_\mu$, $\bar\nu_\mu$, $\nu_\tau$, and $\bar\nu_\tau$ (see e.g., Qian
et al. 1997 for discussions of these reactions in supernovae), we have
solved the system of equations for the production of $^{126}$Te. The bound
on the neutrino fluence obtained from this more accurate
non-perturbative approach is shown as the solid line nearly parallel to the
dot-dashed line in Figure~3. It is more stringent than the approximate
bound in equation~(\ref{f}) because equation~(\ref{y}) underestimates the
production of $^{128}$Sn at earlier times when the neutrino flux is higher.

Previously, Qian et al. (1997) and Haxton et al. (1997) have derived bounds
on the neutrino fluence by considering neutrino-induced neutron spallation
off the $r$-process peak nuclei at $A\sim 130$ and 195 after freeze-out.
They found that the final abundances at $A=124$--126 and 182--187 are
extremely sensitive to the neutrino fluence. By requiring that the nuclei
at $A=124$--126 not be overproduced by neutrino reactions, they put an
upper bound of ${\cal{F}}\lesssim 0.045$ on the neutrino fluence after
freeze-out. This bound is shown as the horizontal solid line in Figure~3.
In addition, Qian et al. (1997) and Haxton et al. (1997) found that for
${\cal{F}}=0.031$, the solar $r$-process abundances at $A=124$--126 can be
produced entirely by neutrino reactions after freeze-out. This neutrino
fluence is shown as the dashed line in Figure~3.
  
\section{Discussion and Conclusions}

Using the Te data in presolar diamonds,
we have derived a new bound on the neutrino fluence and the decay timescale
of the neutrino flux relevant for the supernova $r$-process.
This bound and the previous one of Qian et al.
(1997) and Haxton et al. (1997) define an allowed region shown in Figure~3
for the neutrino fluence ${\cal{F}}$ and the characteristic decay timescale
$\hat\tau$ of the neutrino flux after freeze-out of the $r$-process peak at
$A\sim 130$. In particular, if a neutrino fluence of ${\cal{F}}=0.031$ is
indeed responsible for the production of the solar $r$-process abundances
at $A=124$--126 (Qian et al. 1997; Haxton et al. 1997), 
then the corresponding decay
timescale of the neutrino flux is required to be $\hat\tau\lesssim 0.84$~s. 

The present results depend on a model for retention of 
only the stable isotopes
after implantation of supernova $r$-process products 
into the microdiamonds. The
association of supernovae with the diamond-rich residues found in several
meteorites was first based on the presence of isotopically anomalous Xe
(Xe-HL) that has been attributed to some aspects of the $r$-process (see
the review by Anders \& Zinner 1993; and Clayton 1989). As shown by Richter
et al. (1998), the Te associated with the diamonds has a clear $r$-process
signature and strengthens the possible connection with supernovae. In order
to account for the rather close agreement between
the ${^{128}{\rm Te}}/{^{130}{\rm Te}}$ ratio
in the diamonds and that in the solar system, the
radioactive precursors of $^{128}$Te would have to decay nearly completely
before implantation. 
As noted by Richter et al. (1998),
the decay of $^{128}$Sn (with a half-life of $\tau_{1/2}=59.07$~minutes)
produces $^{128}$Sb predominantly in the short-lived isomeric state
($\tau_{1/2}=10.4$~minutes) rather than
the ground state ($\tau_{1/2}=9.01$~hr). 
It follows that the timescale for retention of the stable isotopes necessary
to produce the observed ${^{128}{\rm Te}}/{^{130}{\rm Te}}$ ratio in the
diamonds must be at least a few hours after $r$-process freeze-out, i.e.,
$t_{\rm ret}\gtrsim 10^4$~s. In addition,
the observed bound ${^{125}{\rm Te}}/{^{130}{\rm Te}}< 0.0018$
limits the decay of $^{125}$Sb before implantation. 
Using the lifetime of $^{125}$Sb, $\tau({^{125}{\rm Sb}})=3.98$~yr,
we obtain 
$t_{\rm ret}<(0.0018/0.159)\tau({^{125}{\rm Sb}})=1.42\times 10^6$~s.
This assumes that the initial $r$-process
yields of the nuclei at $A=125$ and 130 
correspond to ${^{125}{\rm Te}}/{^{130}{\rm Te}}=0.159$,
the ratio of the relevant solar $r$-process abundances.
Thus we obtain $10^4 \lesssim t_{\rm ret} < 1.42\times 10^6$~s.
These bounds on $t_{\rm ret}$ are reasonably consistent with the
retention timescale ($\sim 7.5\times 10^3$~s) calculated by Ott (1996) from
the Xe-H data.

There are several problems that require attention. The first class of
issues is related to the carbon in the diamonds. One of them is that the
${^{13}{\rm C}}/{^{12}{\rm C}}$ ratio in the diamonds is very close to the
solar value. While some variations have been observed (Russell et al.
1991), they are at the level of $\sim 10$\%. In addition, one has to
explain how carbon phases (e.g., diamonds, graphite, SiC) could condense
out of debris from a supernova when the bulk C/O ratio is less than 1. This 
is particularly problematic as the ${^{13}{\rm C}}/{^{12}{\rm C}}$ ratio in
the diamonds in no way reflects the very low ${^{13}{\rm C}}/{^{12}{\rm
C}}$ ratios associated with supernova zones having ${\rm C}/{\rm O}>1$. The
second class of issues is related to the mechanisms of selective
trapping/retention of stable nuclei as compared with 
almost complete loss of radioactive species
and the conditions under which this may take place relative to the
timescales presented here. Many of these issues have previously attracted
attention of other workers. In particular, Clayton (1989) and Clayton et
al. (1995) have proposed models to account for the conditions of formation
of the diamonds, their carbon and nitrogen isotopic composition, and the
implanted Xe. The model of Clayton et al. (1995) for diamond formation
requires a timescale of $\sim 1$~yr and assumes that the essentially solar
${^{13}{\rm C}}/{^{12}{\rm C}}$ ratio is made of a mixture from extremely
$^{12}$C-rich and extremely $^{13}$C-rich supernova zones. The rapid
neutron burst model of Clayton (1989) for the origin of Xe-HL has been
shown to conflict with the Te data by Richter et al. (1998). 

Concerning the issue of the ${^{13}{\rm C}}/{^{12}{\rm C}}$ ratio,
Clayton et al. (1995) recognized that the experimentally analyzed
bulk nanodiamond separates
may represent mixtures of very different components
from many different supernova contributions. 
These workers considered that the
individual nanodiamonds would have 
${^{13}{\rm C}}/{^{12}{\rm C}}$ ratios changing
from their centers to their surfaces. 
Interestingly, with differential thermal and
chemical treatments, it has been demonstrated that there are large
variations ($\sim 50\%$) in the ${^{15}{\rm N}}/{^{14}{\rm N}}$ 
ratio of the bulk diamond
separates (Russell et al. 1991; Verchovsky et al. 1994). 
In addition, the $r$-process Te
and Xe isotopes discussed in this paper correspond to only 1 atom in 
$\sim 10^7$ nanodiamond crystals. The separation of Xe-H from the bulk
Xe-HL by Meshick et al. (1998) is particularly pertinent for the present
discussion. If the $r$-process Te and Xe isotopes
are contained in only a small
subset of the diamonds, then these carriers may have an arbitrary
${^{13}{\rm C}}/{^{12}{\rm C}}$ ratio that is unrelated to the bulk value. 
In this case, the Te and Xe-H carrier grains 
associated with the diamonds could be
pure $^{12}$C and thus represent ejecta from the supernova He/C shell. This
would leave the question of what the bulk diamond separates represent. 
Conceivably, they could be produced in the solar system, thereby explaining 
their observed essentially solar ${^{13}{\rm C}}/{^{12}{\rm C}}$ ratio.

Taking into account some of the issues considered by the previous workers,
we propose the following scenario that may tie together the observations
and try to relate it to plausible supernova processes. We consider either
that a supernova has ${\rm C}/{\rm O}\gtrsim 1$ in its evolved massive
H envelope or that the ejected mass from an interior supernova (He/C) zone 
provides material with ${\rm C}/{\rm O}\gtrsim 1$. 
Upon expansion and cooling after
the supernova explosion, carbon and carbides can be condensed from
the material in the H envelope or the ejected mass
as in the case of AGB stars (cf. Sharp \& Wasserburg 1995;
Bernatowicz et al. 1996). The diamonds or other polymorphs formed in this
way are then dominated by the chemistry and isotopic characteristics
of either the H envelope or 
the ejected material from the interior supernova zones. 

In addition, we consider that the material involved in the $r$-process deep
inside the supernova is blasted out at high velocities. It catches up with
and penetrates the previously ejected material, 
implanting the $r$-process products
into the earlier formed diamond grains (cf. Clayton 1989). We consider that
both the stable and the radioactive nuclei will be implanted. To explain
the absence of $^{125}$Te and $^{126}$Te, we must assume that recoil and
crystallite damage by decay would cause the later loss of essentially all
radioactive species as suggested by Ott (1996). The conditions under which
any of the implanted stable nuclei (particularly the volatile elements)
would be retained suggest a grain temperature for retention of less than
about 800~K. At expansion velocities of $\sim 10^4$~km~s$^{-1}$, the
required cooling could
possibly be achieved for the ejected material within $\sim 10^4$~s,
our lower bound on the retention timescale from the Te data
(i.e., over an expansion distance of $\sim 1$~AU).
Our upper bound of $\sim 10^6$~s 
would allow much more room for cooling
but it would cause a conflict
with the Xe data (Ott 1996). As the size of the diamond crystals is very
small ($\sim 1$~nm), it is possible to achieve grain growth over 
a period of $\sim 10^4$~s
for plausible gas densities and with carbon sticking probabilities of the
order of unity. Loss of radioactive species due to recoil or lattice damage
by decay would take place on a much longer time scale. The typical recoil
energy of the daughter nuclei is $\sim 0.3$--1~eV. This may be enough to
eject them from the diamonds. However, the essentially complete loss of
the implanted radioactive nuclei as compared 
with retention of the implanted stable
nuclei cannot be put on a sufficiently quantitative basis yet. It is also
conceivable that those nuclei with long-lived radioactive precursors
are removed by the severe chemical
processing involved in preparation of the diamond separates due to the
presence of preserved decay-induced defects.

\acknowledgments

This work was supported in part by the U. S.
Department of Energy under Grant No. DE-FG03-88ER-40397,
by NASA under Grant No. NAG5-4076, and by Division
Contribution No. 8524(1005).
Y.-Z. Qian was supported by the David W. Morrisroe 
Fellowship at Caltech.

\clearpage

\figcaption{The time sequence of events relevant for the origin of 
$r$-process Te and Xe isotopes in presolar diamonds. The $r$-process
is considered to take place in the material ejected from the neutron star
produced in a supernova. This material experiences a neutrino flux
that decreases with time. After $r$-process freeze-out, the neutrino
flux is assumed to decay exponentially on a characteristic timescale
$\hat\tau$. A few hours after freeze-out, the $r$-process material
catches up with the previously ejected supernova material, implanting 
$r$-process nuclei
into the earlier formed diamonds. However, only the implanted
stable nuclei will be retained indefinitely when the diamond grains 
become sufficiently cool.}

\figcaption{The three steps involved in neutrino-induced production of 
$^{126}$Te. (a) The production of $^{128}$Sn through $\beta$-decays
after $r$-process freeze-out is shown with the relevant half-lives.
Over the period ($\sim 10$~s) of concern to (b), $^{128}$Sn can be
taken as stable. It decays predominantly to the short-lived isomeric
state of $^{128}$Sb, which in turn decays to $^{128}$Te (shown in the
dotted box). (b) The production of $^{126}$Sb by $\nu_e$ capture
on $^{128}$Sn is shown. It essentially proceeds through 
transitions to the GTGR states in $^{128}$Sb that lie
between the thresholds for emission of two and three neutrons
(indicated as the dotted lines).
Due to angular momentum selection rules, 
$^{126}$Sb is produced mainly in its isomeric state rather 
than its high-spin ground state.
(c) The production of $^{126}$Te
through the $\beta$-decay of $^{126}$Sb is shown. 
The $\beta$-decay from the
isomeric state dominates the production [see (b)].
For comparison, the production of $^{126}$Te 
through the $\beta$-decay of $^{126}$Sn on a much longer timescale is
shown in the dotted box. Note that the nuclear energy levels in Fig.~1
are for illustrative purpose only and are not to scale.}

\figcaption{Bounds on the neutrino fluence ${\cal{F}}$ and the 
characteristic decay timescale $\hat\tau$ of the neutrino flux
after freeze-out of the $r$-process peak at $A\sim 130$. 
The dot-dashed line shows the approximate bound 
(eq.~[{\protect\ref{f}}]) derived
from the data of Richter et al. (1998) on Te isotopes in presolar
diamonds. The solid line nearly parallel to it is from a more accurate
calculation. The horizontal solid line is from Qian et al. (1997)
and Haxton et al. (1997). The dashed line shows the neutrino fluence
that may be responsible for the production of the solar $r$-process
abundances at $A=124$--126. The Te data in presolar diamonds require
that such a fluence be provided by a neutrino flux with 
$\hat\tau\lesssim 0.84$~s.}

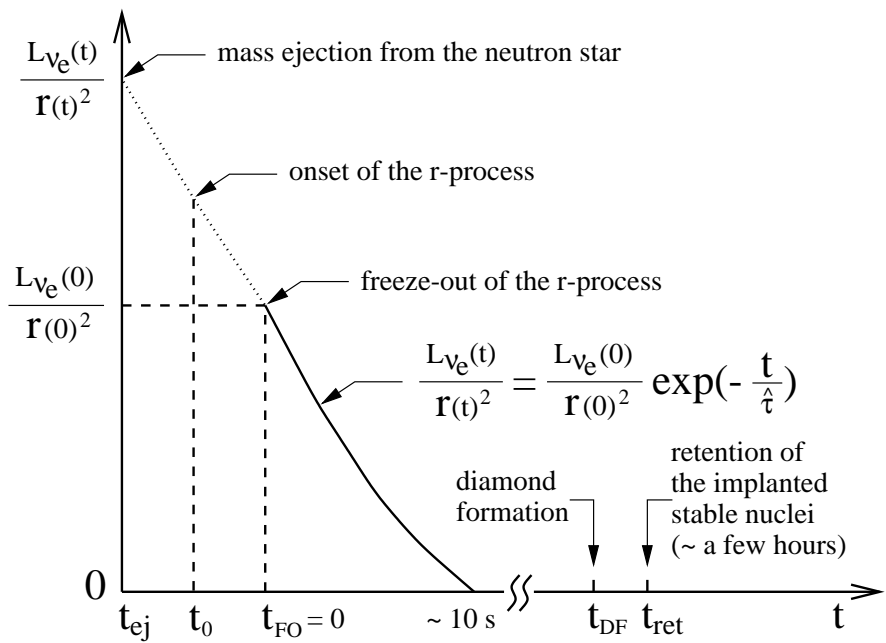
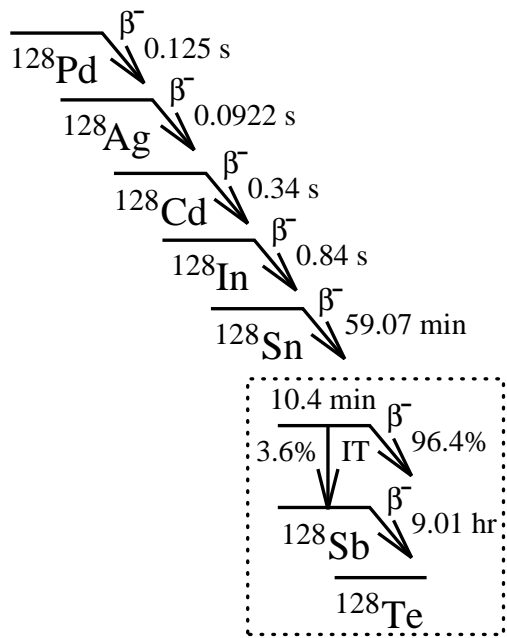
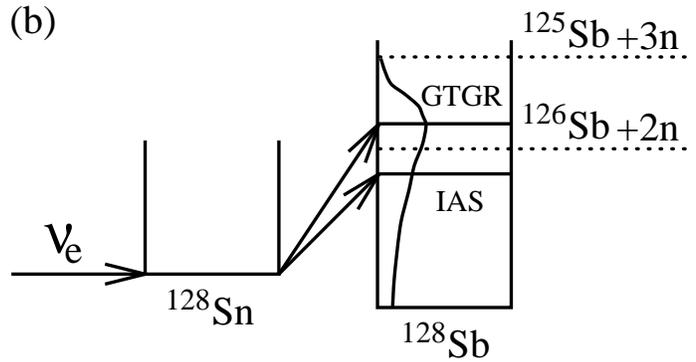
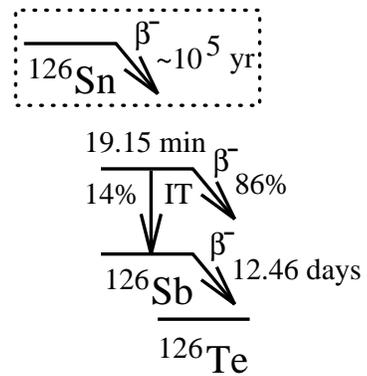
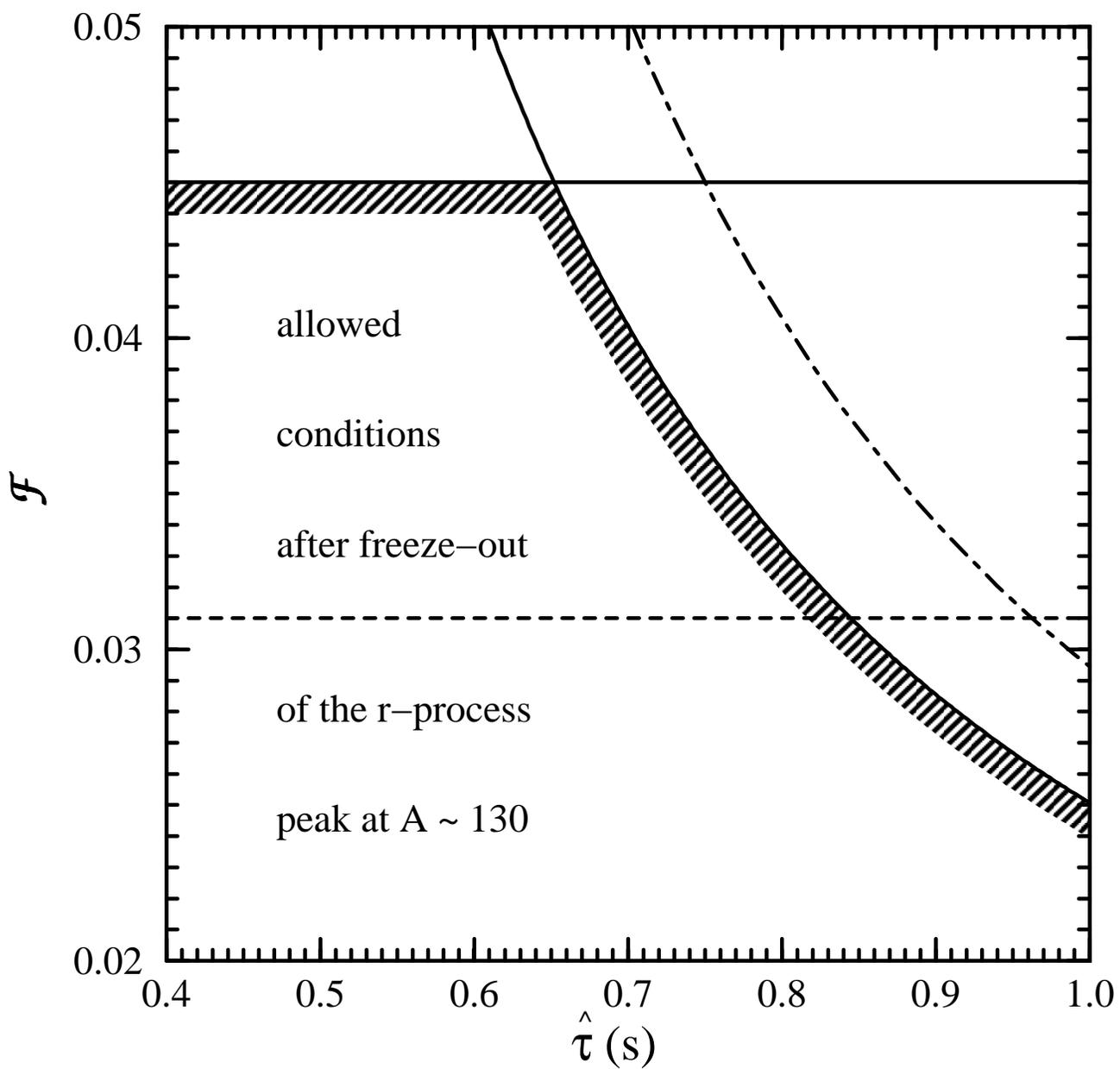
\end{document}